\documentclass[twocolumn,showpacs,amsmath,amssymb]{revtex4}

\usepackage{graphicx}
\usepackage{dcolumn}
\usepackage{bm}

\begin{document}

\title{Detrended fluctuation analysis on the correlations of complex networks \\
under attack and repair strategy}

\author{L.P. Chi$^\dag$, C.B. Yang, K. Ma, and X. Cai}
\affiliation{Institute of Particle Physics,
HuaZhong (Central China) Normal University, Wuhan 430079, China \\
$^\dag$Email of the Corresponding Author: chilp@iopp.ccnu.edu.cn}

\date{\today}

\begin{abstract}

We analyze the correlation properties of the Erd\"{o}s-R\'{e}nyi
random graph (RG) and the Barab\'{a}si-Albert scale-free network
(SF) under the attack and repair strategy with detrended
fluctuation analysis (DFA). The maximum degree $k_{max}$,
representing the local property of the system, shows similar
scaling behaviors for random graphs and scale-free networks. The
fluctuations are quite random at short time scales but display
strong anticorrelation at longer time scales under the same system
size $N$ and different repair probability $p_{re}$. The average
degree $\langle k \rangle$, revealing the statistical property of
the system, exhibits completely different scaling behaviors for
random graphs and scale-free networks. Random graphs display
long-range power-law correlations. Scale-free networks are
uncorrelated at short time scales; while anticorrelated at longer
time scales and the anticorrelation becoming stronger with the
increase of $p_{re}$.

\end{abstract}



\maketitle

\section{Introduction}

Complex networks are an essential part of modern society. Many
social, biological, transportation and communication systems can
be cast into the form of complex networks, a set of nodes joined
together by links indicating interactions. \cite{s1,s2,s3,s4}
Recently enormous interest has been devoted to the study of
complex networks. Among various existing models of complex
networks, two different model topologies, the Erd\"{o}s-R\'{e}nyi
model \cite{s5} of the random graph (RG) and the
Barab\'{a}si-Albert model \cite{s6} of the scale-free network
(SF), have been widely studied. \cite{s7,s8,s9}

The random graph is constructed starting from an initial condition
of $N$ nodes and no edges. Then edges are added between pairs of
randomly selected nodes with connection probability $p_{rg}$. The
number of edges connected to any particular node is called the
degree $k$ of that node. The average degree of the random graph is
$\langle k \rangle =  Np_{rg}$. The scale-free network puts the
emphasis on the network dynamics and is constructed with the
algorithm of \textit{growth} and \textit{preferential attachment}.
According to the scale-free model, the network grows over time by
the addition of new nodes and links. A node newly added to the
network randomly selects $m$ other nodes to establish new links,
with a selection probability that increases with the number of
links of the selected node. One of the most relevant is given by
power-law degree distribution $P(k)$, defined as the probability
that a randomly chosen node has degree $k$, $P(k) \sim
k^{-\gamma}$.

Recently the security of complex networks to the random failures
or intentional attacks has attracted a great deal of attention.
\cite{s10,s11,s12,s13,s14,s15} The random failure is simulated as
the deletion of network nodes or links chosen at random, while
intentional attack as the targeted removal of a specific class of
nodes or links. In order to protect existing networks and design
robust networks, we introduced the attack and repair strategy in
Ref. [16], which will be described briefly in the following
section. In this paper, we will study the correlations for random
graphs and scale-free networks under the attack and repair
strategy with the detrended fluctuation analysis (DFA)
\cite{s17,s18} and investigate the intrinsic dynamics in both
networks. The advantages of DFA over many methods are that it
permits the detection of long-range correlations embedded in
seemingly nonstationary time series, and also avoids the spurious
detection of apparent long-range correlations that are an artifact
of nonstationary. \cite{s19,s20,s21,s22}

The paper is organized as follows. In Section II we review the
algorithm of the DFA method. In Section III, we apply the DFA
method to investigate the correlations of the maximum degree
$k_{max}$ and the average degree $\langle k \rangle$ for random
graphs and scale-free networks. In last section we present our
brief conclusions.

\section{DFA}

Detrended fluctuation analysis (DFA) is a well-established method
for determining the scaling behavior of noisy data in the presence
of trends without knowing their origin and shape. Briefly, the DFA
method involves the following steps.

\begin{enumerate}

\item[(1)]
We consider a noisy time series $u(i)$ ($i=1,..,N$) and $N$ is the
length of the series. We integrate the time series $u(i)$ and
subtract the mean $\langle u \rangle$,

\begin{equation}
y(k) = \sum\limits_{i=1}^{k}[u(i) - \langle u \rangle],
\end{equation}

where

\begin{equation}
\langle u\rangle = \frac{1}{N}\sum_{i=1}^{N}u(i).
\end{equation}

\item[(2)]
We divide the integrated signal $y(k)$ into non-overlapping boxes
of equal size $n$ (scale of analysis).

\item[(3)]
In each box of size $n$, we fit the integrated time series $y(k)$
by using a polynomial function, $y_{fit}(k)$, which is called the
local trend. For order-$\ell$ DFA (DFA-1 if $\ell=1$, DFA-2 if
$\ell=2$ etc.), $\ell$-order polynomial function is applied for
the fitting.

\item[(4)]
We detrend the integrated time series $y(k)$ by subtracting the
local trend $y_{fit}(k)$ in each box, and we calculate the
detrended fluctuation function

\begin{equation}
Y(k) = y(k)-y_{fit}(k).
\end{equation}

\item[(5)]
For a given box size $n$, we calculate the root-mean-square (rms)
fluctuation

\begin{equation}
F(n) = \sqrt{\frac{1}{N}\sum_{k=1}^{N}[Y(k)]^2}
\end{equation}

and repeat the above computation for different box sizes $n$
(different scales) to provide a relationship between $F(n)$ and
$n$.

\end{enumerate}

A power-law relation $F(n) \sim n^{\alpha}$ between $F(n)$ and the
box size $n$ indicates the presence of scaling. The parameter
$\alpha$, called the scaling exponent or correlation exponent,
represents the correlation properties of the signal. If
$\alpha=0.5$, there is no correlation and the signal is an
uncorrelated signal (white noise); if $\alpha < 0.5$, the signal
is anticorrelated; if $\alpha >0.5$, there are positive
correlations in the signal.

\section{DFA on correlations of RG and SF}

First we will give a brief description about the attack and repair
strategy. We start by constructing a network according to the
algorithms of Erd\"{o}s-R\'{e}nyi random graph model (RG) or
Barab\'{a}si-Albert scale-free model (SF). The dynamics of the
attack and repair strategy is defined in terms of the following
two operations:

\begin{itemize}

\item
\textit{Attack}: Find a node with the maximum degree $k_{max}$ and
remove all its links. (If several nodes happen to have the same
highest degree of connection, we randomly choose one of them.)

\item
\textit{Repair}: Reconnect this node with the other nodes in the
network with repair probability $p_{re}$.

\end{itemize}

Then, the evolution comes into the next Monte Carlo time step. At
each Monte Carlo time step $s$, we record the maximum degree
$k_{max}(s)$ and the average degree $\langle k \rangle(s)$ of the
system. The reason we choose $k_{max}$ and $\langle k \rangle$ to
analyze is that they represent the local and statistical features
of the system. Since the time step should be large enough to
enable the system to reach stationary state, it is chosen to be
1.5 million in our simulation. To reduce the effect of fluctuation
on calculated results, for every system with size $N$, the
calculated results $k_{max}$ and $\langle k \rangle$ are averaged
over 10 independent runs for each network realization.

\subsection{DFA on the maximum degree $k_{max}$}

According to the attack and repair strategy, it can be easily seen
that the network has a tendency to decrease the maximum number of
connections among the nodes at a long time scale, because the
nodes with highest connections will be damaged and reconnected
randomly.

\begin{figure}
\begin{center}
\includegraphics[height=8cm,width=8cm]{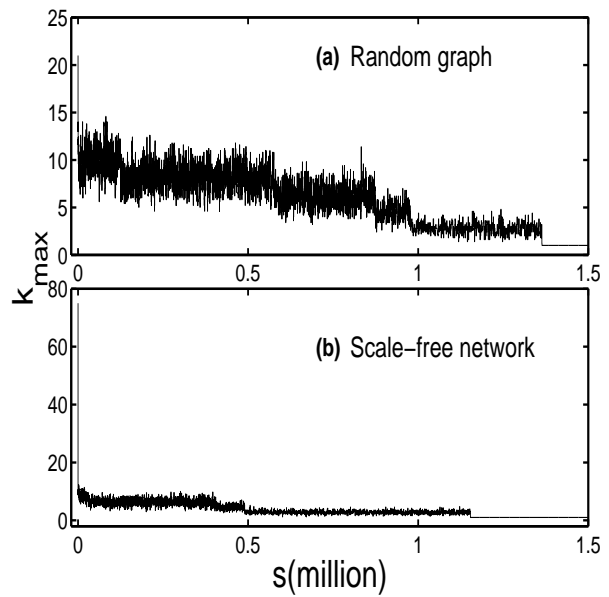}
\caption{The maximum degree $k_{max}$ for (a) random graph and (b)
scale-free network under $N=1000$ and $p_{re}=0.01$.}
\end{center}
\end{figure}

In Fig. 1, we plot the maximum degree $k_{max}$ versus time step
$s$ for RG and SF with $N=1000$ nodes and repair probability
$p_{re}=0.01$. It should be illustrated here that, in order to
reduce the number of parameters in the model, we set the repair
probability $p_{re}$ the same as the connection probability
$p_{rg}$ in RG. Fig. 1 shows that $k_{max}$ of RG decreases slowly
and fluctuates widely, while $k_{max}$ of SF decreases very
steeply at the beginning and then very slowly with small
fluctuations. The different behaviors of $k_{max}$ for RG and SF
are rooted in their different topological structures. RG has
homogeneous Poissonian degree distribution: all nodes have
approximately the same number of links. The removal of each node
causes the same amount of damage. SF is an extremely inhomogeneous
network: because the power-law distribution implies that the
minority of nodes have highly connected nodes. The removal of
these nodes drastically decreases the degrees of the network.

\begin{figure}
\begin{center}
\includegraphics[height=8cm,width=8cm]{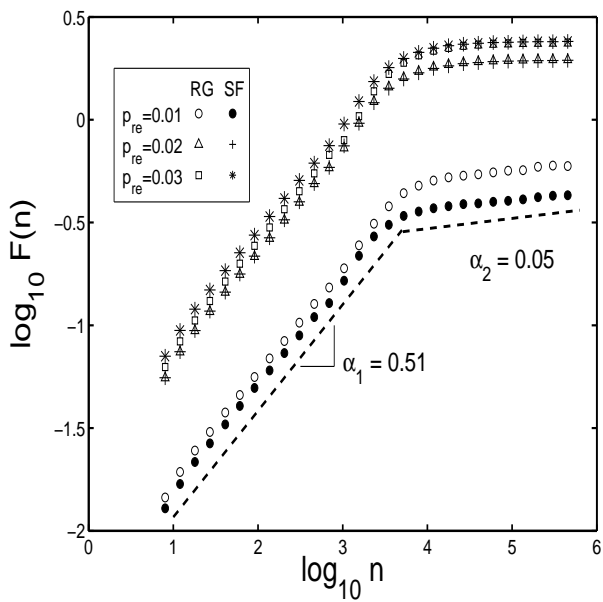}
\caption{Scaling behavior of the root-mean-square (rms)
fluctuation function $F(n)$ as a function of the scale $n$, for
series $k_{max}$ of random graphs (RG) and scale-free networks
(SF) under $N=1000$ and $p_{re}=$ 0.01, 0.02, 0.03, respectively.
The dashed line represents the fit of the slope.}
\end{center}
\end{figure}

In Fig.2 we present results from the DFA-3 analysis for the
maximum degree $k_{max}$ of RG and SF. The system size is $N=1000$
nodes and the repair probability $p_{re}$ are 0.01, 0.02, 0.03,
respectively. In this double logarithm graph, the root-mean-square
(rms) fluctuation function $F(n)$ as a function of the scale $n$,
exhibit two distinct scaling regimes with the crossover at about
$n=10^4$,

\begin{equation}
F(n)\sim \left\{\begin{array}{ll} n^{\alpha_1}, & {n < 10^4}; \\
n^{\alpha_2}, & {n > 10^4}.
\end{array}\right.
\end{equation}

\noindent For $n<10^4$, the scaling exponent $\alpha_1$ is about
0.51; while for $n>10^4$, the exponent $\alpha_2$ is about 0.05.
Our DFA results suggest that the fluctuation is quite random
(close to white noise) for short time scales. As the time scale
becomes larger, the fluctuation displays strong anticorrelation,
that is, a large value (compared to the average) of $k_{max}$ is
almost always followed by a small value. With the increase of the
repair probability $p_{re}$, we find that there is a vertical
shift of the rms fluctuation function $F(n)$ to larger values and
the shift between $p_{re}=0.01$ and 0.02 is very big. In addition,
the similar shape between $F(n)$ and $n$ is observed for both RG
and SF under different $p_{re}$, which illustrates that the time
series of $k_{max}$ for RG and SF have the similar correlation
properties.

\subsection{DFA on the average degree $\langle k \rangle$}

\begin{figure}
\begin{center}
\includegraphics[height=8cm,width=8cm]{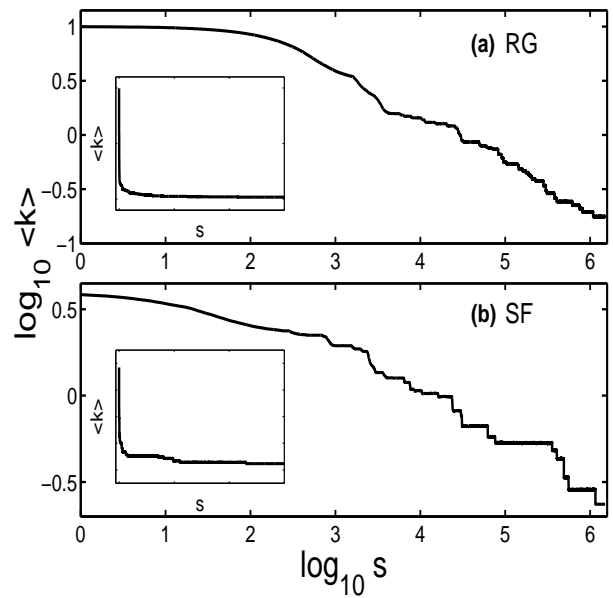}
\caption{The average degree $\langle k \rangle$ for (a) random
graph and (b) scale-free network under $N=1000$ and
$p_{re}=0.01$.}
\end{center}
\end{figure}

The average degree $\langle k \rangle$ is a statistical
topological property of complex networks. The time series of
$\langle k \rangle$ reflect the general changes of RG and SF under
the evolution of attack and repair. We find that $\langle k
\rangle$ of both RG and SF decay exponentially with small
fluctuations as shown in Fig. 3, under the system size $N=1000$
and the repair probability $p_{re}=0.01$. The inset panel of Fig.
3 gives the series of $\langle k \rangle$ in linear coordinates,
showing that $\langle k \rangle$ decreases very steeply at first
and then becomes steadily. In fact, the curve of $\langle k
\rangle$, to a large degree, depends on the repair probability
$p_{re}$. The bigger the repair probability, the smoother the
curve will be. Here we choose $p_{re}=0.01$ to analyze,
considering to have a comparison with the results of $k_{max}$.

We next investigate the correlation properties of series $\langle
k \rangle$. Fig. 4(a) shows the results of applying DFA-3 to the
series $\langle k \rangle$ of RG under $N=1000$ and $p_{re}=$0.01,
0.02, 0.03, respectively. In the log-log coordinates, the rms
fluctuation function $F(n)$ and the scale $n$ has a power-law
dependence with scaling exponent $\alpha = 0.64$, which indicates
the presence of the positive correlations and the weak long-range
(long-memory) power-law correlations. The results suggest that a
value of $\langle k \rangle$ at a given time depends on values of
long past in a power-law fashion. We also note that $F(n)$ shifts
downwards slightly with the increase of $p_{re}$, which is
oppositive to that of $k_{max}$ in Fig. 2. Besides, the big shift
between $p_{re}=0.01$ and 0.02 vanishes here.

Fig. 4(b) reports the DFA-3 results of time series $\langle k
\rangle$ for SF under $N=1000$ and $p_{re}=$0.01, 0.02, 0.03,
correspondingly. The two-segment power-law relationship is
observed with the crossover at about $n=10^3$,

\begin{equation}
F(n)\sim \left\{\begin{array}{ll} n^{\alpha_1}, & {n < 10^3}; \\
n^{\alpha_2}, & {n > 10^3}.
\end{array}\right.
\end{equation}

\noindent For short time scales ($n < 10^3$), the rms fluctuation
function $F(n)$ has the similar scaling exponent with
$\alpha_1=0.55$ for different repair probability $p_{re}$. For
longer time scales ($n
> 10^3$), the exponents $\alpha_2$ of $p_{re}=0.01$, 0.02, 0.03 correspond to 0.54, 0.36,
0.23. It shows that the rms fluctuation function $F(n)$ decreases
and the series $\langle k \rangle$ becomes strong anticorrelated
with increasing $p_{re}$ for longer time scales.

\begin{figure}
\begin{center}
\includegraphics[height=8cm,width=8cm]{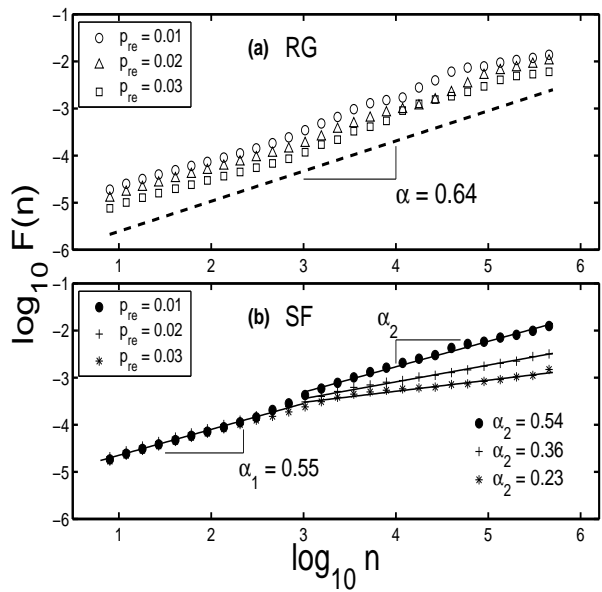}
\caption{Scaling behavior of the root-mean-square (rms)
fluctuation function $F(n)$ as a function of the scale $n$, for
series $\langle k \rangle$ of (a) random graphs (RG) and (b)
scale-free networks (SF), under $N=1000$ and $p_{re}=$ 0.01, 0.02,
0.03, respectively. The dashed line for RG and the solid line for
SF represent the fit of the slopes.}
\end{center}
\end{figure}

Fig. 4(a) and 4(b) demonstrate that the rms fluctuation function
$F(n)$ of series $\langle k \rangle$ exhibits different scaling
behaviors for RG and SF. The positive long-lange power-law
correlations of RG are probably due to the strong restriction,
defined as the connection probability equals the repair
probability, $p_{rg}=p_{re}$. For scale-free networks, the repair
probability $p_{re}$ has a great effect on the power-law degree
distribution of SF in small ranges. The randomness are dominated
at small time scales, characterized by uncorrelated series. While
at large time scales, the power-law degree distributions are
dominated, characterized by the anticorrelation.

Further, Fig. 2 and Fig. 4 represent the DFA results of series
$k_{max}$ and $\langle k \rangle$. We note that, with the increase
of $p_{re}$, the rms fluctuation function $F(n)$ increases in
$k_{max}$ but decreases in $\langle k \rangle$. It is the
consequence that more random connections are added with increasing
$p_{re}$. According to the local property of $k_{max}$, the
fluctuation of the system increases with added randomness. While
according to the statistical property of $\langle k \rangle$, the
fluctuation of the system decreases since the added connections
balanced the degrees of the whole network.

\section{Conclusions}

In summary, we use DFA method to analyze the correlation
properties of random graphs and scale-free networks under the
attack and repair strategy. According to the analysis of the
maximum degree $k_{max}$ representing the local property of the
system, we find that both random graphs and scale-free networks
exhibit the similar scaling behaviors, which suggest that the
fluctuations are quite random at short time scales but strong
anticorrelated at longer time scales. With the analysis of the
average degree $\langle k \rangle$ representing the statistical
property of the system, we observe that random graphs and
scale-free networks display completely different scaling
behaviors. Random graphs show long-range power-law correlations;
while scale-free networks are uncorrelated at short time scales
but display anticorrelations at longer time scales. With the
increase of the repair probability $p_{re}$, the apparent vertical
shifts of the rms fluctuation function $F(n)$ for $k_{max}$ and
$\langle k \rangle$ are due to the effects of the increase of
random connections to the networks.

\section*{Acknowledgments}

This work was supported in part by the National Natural Science
Foundation of China under grant Nos. 70271067, 70401020 and by the
Ministry of Education of China under grant No. 03113.

\end{document}